\documentclass[fleqn,10pt]{wlscirep}
\usepackage[utf8]{inputenc}
\usepackage[T1]{fontenc}
\usepackage{bm}
\title{Solid immersion metalens for directional single molecule emission with high collection efficiency}

\author[1]{Zhiheng Li}
\author[1]{Zequan Chen}
\author[1]{Rupert F. Oulton}
\author[1,*]{Ming Fu}
\affil[1]{The Blackett Laboratory, Department of Physics, Imperial College London, London SW7 2AZ, United Kingdom}
\affil[*]{e-mail: ming.fu@imperial.ac.uk}

\begin{abstract}
We present simulations of an efficient high numerical aperture solid immersion metalens concept for fluorescence microscopy. The technique exploits the preferential emission of interfacial dipoles into a high refractive index substrate combined with a metalens and a conventional tube lens for imaging them. We have thus simulated dipole emission and an all-dielectric metasurface on opposite sides of a high refractive index substrate. Our calculations predict dipole collection efficiencies of up to $87$ percent. The simulated beam propagation through the imaging system shows excellent performance along the optical axis, with aberrations accumulating with increasing field of view. These aberrations can be controlled by using a metasurface with an optimized non-hyperbolic phase profile. The high collection efficiency of dipole emission suggests this compact solid immersion lens would be effective for fluorescence imaging including single fluorescent centres for quantum optical application.
\end{abstract}
\begin{document}

\flushbottom
\maketitle

\thispagestyle{empty}

\section*{Introduction}
Isolated fluorescent centres are widely used as sources for super-resolution imaging \cite{betzig_imaging_2006,zhuang_nano-imaging_2009} as well as quantum optics \cite{lounis_single_2000}, where both applications rely on a high optical collection efficiency.\cite{Lee2011} Since the radiation of dipole emitters is omnidirectional, numerous elaborate techniques have been devised to make the radiation more directional \cite{sciencenanoantenna,nanoantenna2,microcavity1,microcavity2,photonicnanowire1,photonicnanowire2}. However, even the simple act of placing a dipole emitter at the interface between two dielectric media can have a profound impact on its radiation pattern.\cite{Lee2011,flourescencemicroscopy,NVcentre} The interface of air with a substrate of refractive index, $n_S$, can even be viewed as a dielectric antenna, since it promotes dominant emission into the substrate due to a higher substrate density of states, which scales with $n_S^3$. When suitably optimised for dipole emitters orientated perpendicular to the interface, Lee and colleagues achieved near-unity collection efficiency.\cite{Lee2011} However, the resulting high angle annular-shaped emission still required a high numerical aperture (NA) solid immersion imaging system. In this work, we show that high NA radiation patterns from interfacial dipoles may be collected through the substrate using a metalens to collimate the high angle annular emission patterns, and subsequently imaged using a tube lens. The approach proves to be both efficient and effective for collecting emission independently of dipole orientation. 

\section*{Results and Discussion}
\begin{figure}[t]
\centering
\includegraphics[width=0.5\linewidth]{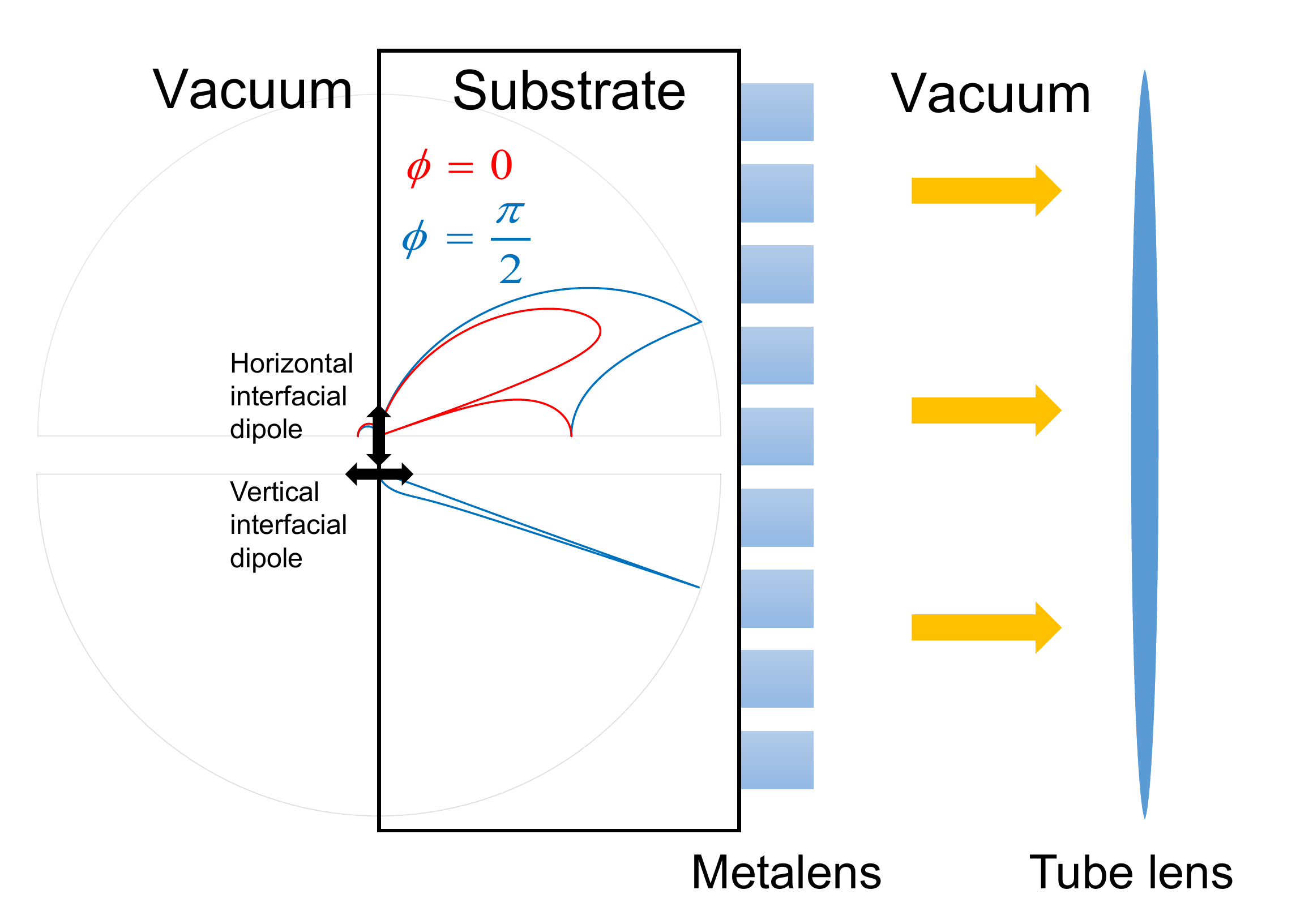}
\caption{The schematic design of the high NA interfacial dipole solid immersion metalens. The interfacial dipole radiation patterns are shown for a substrate refractive index of $n_S=3$.}
\label{fig:figure1}
\end{figure}

 Metasurfaces are two-dimensional metamaterials composed of nanostructured building blocks that enable a local control of transmission phase. For example, a hyperbolic spatial phase profile metasurface defines a simple metalens.\cite{metalens1,metalens2,metalens3,metalens4} The schematic design of the combined dielectric antenna and metalens is presented in Figure \ref{fig:figure1}. The necessary $2\pi$ phase control over a metasurfaces is often achieved by a geometric phase associated with circularly polarised light interacting with spatially varying anisotropic building blocks\cite{metasurface1,anisotropic1,anisotropic2}. This provides robust wavelength-independent phase control, but its dependence on polarization limits its use for imaging of incoherent light. Alternatively, metasurfaces composed of isotropic building blocks are polarisation-independent, but suffer a wavelength-dependence, which must be managed.\cite{metalens1,isotropic1,isotropic2} Since randomly oriented fluorescent dipoles typically emit over a narrow spectral range, but with undefined polarization, we employ the latter approach to fabricate our metalens.\cite{metalens1} In the following, we have assumed dipole emission at a wavelength 0.6 $\mu{m}$ on a substrate with thickness 100 $\mu{m}$ or 250 $\mu{m}$, and different refractive indices.  

\begin{figure}[b]
\centering
\includegraphics[width=0.8\linewidth]{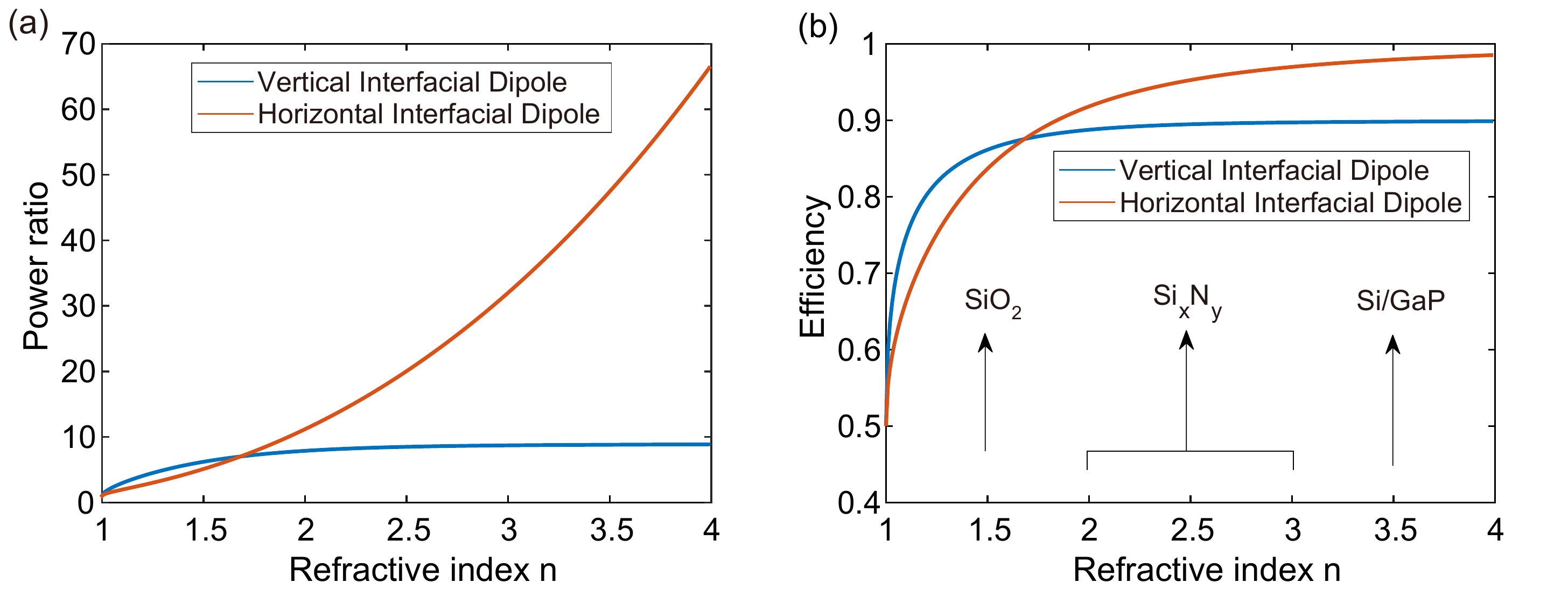}
\caption{(a) The power ratios and (b) efficiencies as a function of the refractive index for vertical and horizontal interfacial dipoles. Some commonly used dielectric substrate materials including silicon dioxide, silicon rich nitride, silicon and gallium phosphide are labeled.}
\label{fig:figure2}
\end{figure}

We first discuss interfacial dipole radiation. It is well-known that a dipole in a homogeneous medium has a doughnut-shaped radiation pattern. According to Fermi's golden rule, the radiative decay rate of a dipole emitter depends on the density of states, which is proportional to the cube of the refractive index for a homogeneous medium.\cite{ncubicdecay} Therefore, placing a dipole emitter at a dielectric interface preferentially directs emission into the high-index substrate.\cite{Lee2011} The problem of interfacial dipoles, solved by Engheta\cite{engheta}, shows that this logic works well for dipoles oriented parallel to the interface; but a rigorous description is required to treat vertical dipoles and to identify the complicated radiation patterns, as shown in Figure \ref{fig:figure1}. We define a dipole oriented perpendicular to the interface as a vertical interfacial dipole, while the parallel orientations are considered horizontal interfacial dipoles. For the vertical interfacial dipole, the radiation into the dielectric half space is rotationally symmetric and the radiation forms a narrow peak near the critical angle $\theta_C$, where $\theta_C=\sin^{-1}{\left(1/n\right)}$. Meanwhile, the horizontal dipole's radiation pattern is not rotationally symmetric, with peak emission at an azimuthal angle of $\phi=\pi/2$, and with radiation dominant at angles above the critical angle. Although the interfacial dipole radiation patterns are drastically distorted, the phase fronts for both dipole orientations remain spherical and this enables the re-focusing of the radiation patterns to form an image using a simple metalens. 

To estimate how efficiently the substrate collects dipole radiation, we have simulated the power emitted into both vacuum ($P_{vac}$) and substrate ($P_{sub}$) to evaluate both the power ratio, $P_{sub}/P_{vac}$ and the efficiency, $P_{sub}/(P_{sub}+P_{vac})$ as a function of the refractive index of the substrate for both dipole orientations. Figure \ref{fig:figure2} shows power ratios for both dipole orientations increase with substrate index. Notably, the horizontal dipole follows the expected $n_s^3$ power law.\cite{ncubicdecay} Both the vertical and horizontal interfacial dipoles' efficiencies saturate rapidly, so extremely high refractive indices are not necessary. For example, we find vertical and horizontal dipole substrate emission efficiencies of $0.8875$ and $0.918$ respectively when the substrate index is just $2$, a value attainable with many transparent materials. 

Although close to 90 percent of a dipole's energy is directed into the substrate, a large proportion of the radiation will only be collected at high NA. The advantage of metalenses are their capability to operate at high NA in a planar configuration.\cite{metalens1} To demonstrate the importance of achieving a high NA, we have compared the amount of dipole energy collected into two NAs of $1.44$ and $2.00$ respectively, as presented in Table \ref{tab:metalens_eff}. In Table \ref{tab:metalens_eff} the efficiency is defined as the power collected by the metalens divided by the total power emitted from the dipole. The efficiencies for the slightly reduced NA drop to almost half. Let us now consider the construction of the metalens and its transmission efficiency.

\begin{table}[t]
\centering
\caption{\bf The efficiencies collected by the metalens with two different NAs for the vertical/horizontal interfacial dipole. The thickness of the substrate is set to be 100 $\mu{m}$}
\begin{tabular}{ccccc}
\hline
Numerical aperture  & Efficiency & Efficiency \\
  & Vertical dipole & Horizontal dipole  \\
\hline
$1.44$ & $0.55$ & $0.45$  \\
$2.00$ & $0.85$ & $0.87$ \\
\hline
\end{tabular}
\label{tab:metalens_eff}
\end{table}

The chosen metalens design is an array of cylindrical nanopillars arranged on a square lattice. The isotropic geometry of the nanopillars results in a polarization-insensitive transmission. The metalens should function as a solid immersion lens and convert the spherical wave emission from interfacial dipoles into a collimated beam. Consequently, the spatial phase profile of the metalens must be hyperbolic. If we assume the beam propagation direction is $z$ and the metalens is placed in the $z=0$ plane, then the phase profile of the solid-immersion metalens is given by:
\begin{equation}
    \phi_{target}=-\frac{2\pi}{\lambda}n\left[\sqrt{x^2+y^2+f^2}-f\right]
\end{equation}
\noindent where $\lambda$ is the wavelength, $n$ is the refractive index of the substrate, $x$ and $y$ are the coordinates in the metalens plane, and $f$ is the focal length of the metalens. The focal length is set equal to the thickness of the substrate, which is 100 $\mu{m}$. Here, we have set the nano-pillars to be TiO$_2$ with refractive index of 2.4.\cite{TiO2_broadband_2016} The substrate is set to be AlN with refractive index of 2.15.\cite{AlN_refraction_1966} The refractive index of the pillars should be higher than the substrate to provide the $2\pi$ phase coverage with fabrication limited nano-pillar sizes. The height of the nano-pillars is 0.6 $\mu{m}$ and the square unit cell size is 0.35 $\mu{m}$, as shown in the inset of Figure 3a. The sub-wavelength periodic grating ensures that higher order diffracted waves are evanescent and almost all the energy is directed into the zeroth order. 

\begin{figure}[t]
\centering
\includegraphics[width=0.8\linewidth]{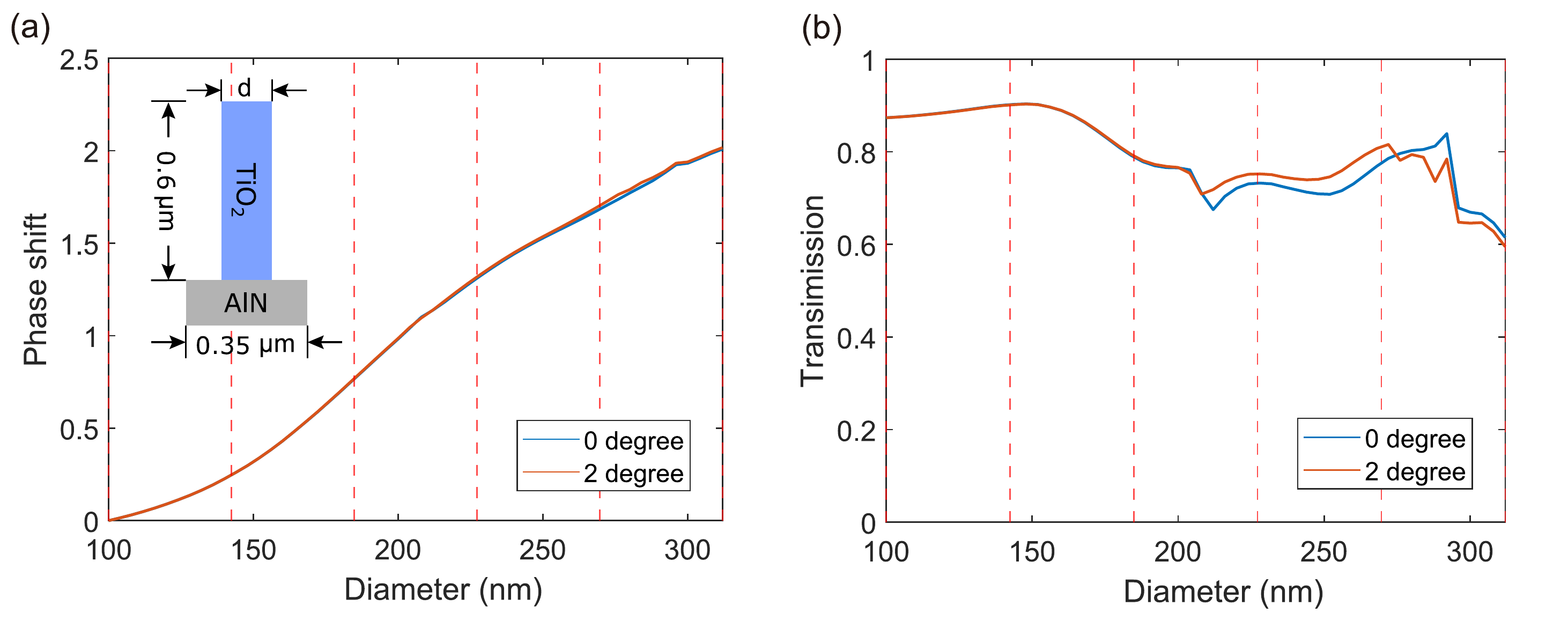}
\caption{(a) The phase shift and (b) transmission as a function of the nanopillar diameter at the field angles of $0$ and $2$ degrees. The inset shows the scheme of one unit cell, with $n_{TiO_2}$ = 2.4 and $n_{AlN}$ = 2.15. $d$ is the nanopillar diameter.}
\label{fig:figure3}
\end{figure}

We now vary the nano-pillar diameters range from 0.1 $\mu{m}$ to 0.312 $\mu{m}$. Therefore, the duty cycle (ratio of nano-pillar diameter to the unit cell size) is from $0.29$ to $0.89$. In order to investigate the effect of the phase and transmission relations on the field of view, we performed finite difference time domain (FDTD) simulations to characterize the phase shift and transmission as a function of the duty cycle at normal incidence and at the field angle of 2 degrees. As shown in Figure 3a, the phase gradient  for a larger incident angle is nearly the same as that at normal incidence. Similarly, as demonstrated in Figure 3b there are minor differences in transmission for the two field angles. The total transmission of the metalens for the field angles of 0 and 2 degrees are around $78.7$\% and $79.8$\% respectively. From the simulation above, we can conclude that the variation of the phase and transmission for different field angles does not limit the field of view beyond conventional lens aberrations. The range of the duty cycle should provide $2\pi$ phase coverage with as high a transmission as possible. Noticeable phase fluctuations and transmission dips are apparent in Fig. 3a and 3b, due to nano-pillar resonances.\cite{isotropic1,zhan2016low} These could be narrowed or even eliminated by increasing the nanopillar height or reducing the unit cell size, but these operations lead to a larger aspect ratios which would be more challenging to fabricate. In order to keep an acceptable aspect ratio and avoid resonances, we chose six different duty cycles to avoid them. The chosen diameters are indicated with dashed line in Figure \ref{fig:figure3}. Given the phase relation at normal incidence presented in Figure 3a, we can construct the metalens by fitting the spatial phase profile to the duty cycle. 

\begin{figure}[t]
\centering
\includegraphics[width=0.65\linewidth]{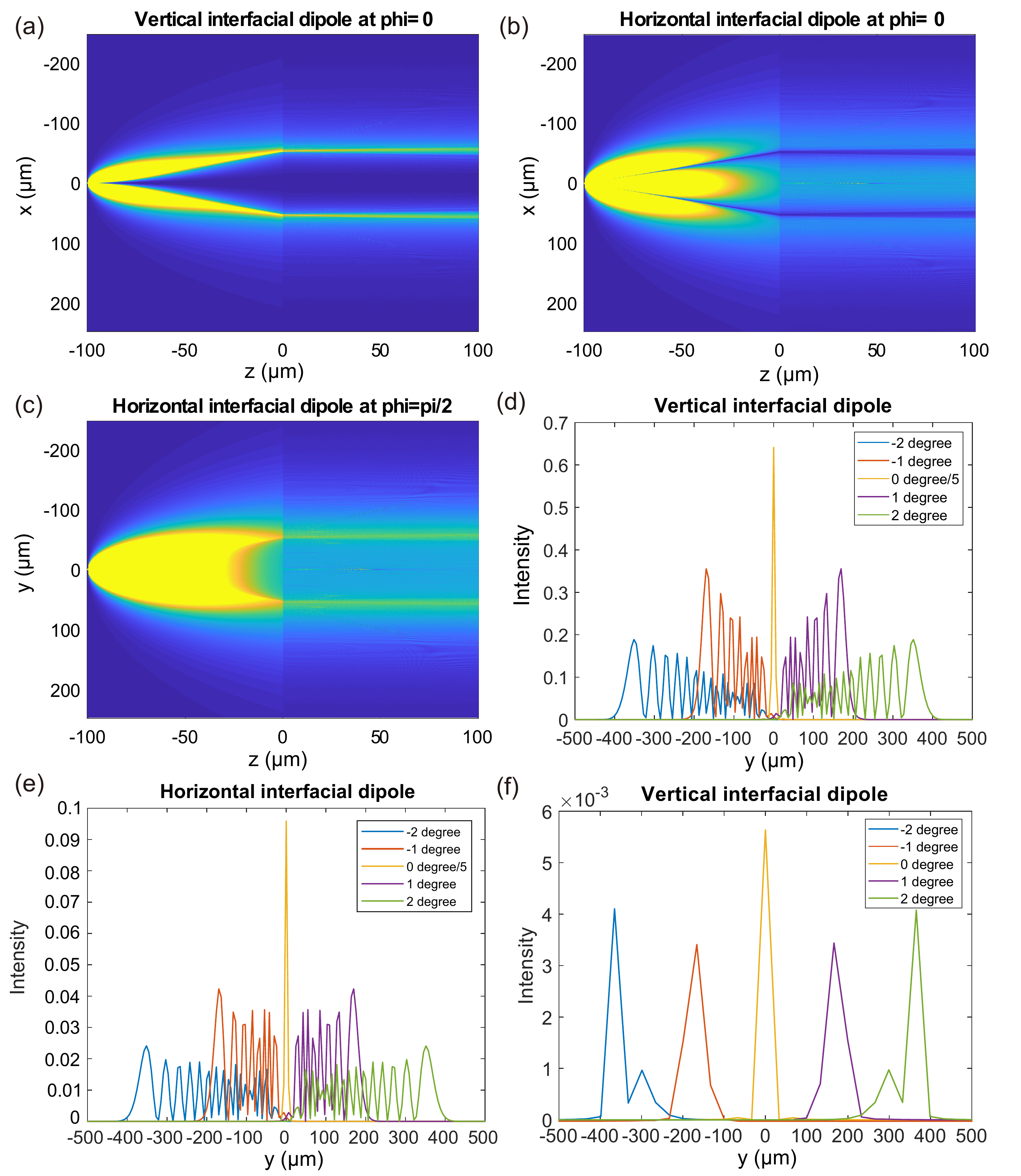}
\caption{The beam propagation for (a) the vertical interfacial dipole at $\phi=0$; (b) the horizontal interfacial dipole at $\phi=0$ and (c) the horizontal interfacial dipole $\phi=\frac{\pi}{2}$. The intensity distributions as a function of $y$ for (d) the vertical interfacial dipoles and (e) the horizontal interfacial dipoles at the field angles of $\pm{2}$ degrees, $\pm{1}$ degree and $0$ degree with the metalens NA of $2.00$. (f) The intensity distribution as a function of $y$ for the vertical interfacial dipoles at the field angles of $\pm{2}$ degrees, $\pm{1}$ degree and $0$ degree with the metalens NA of $0.89$.}
\label{fig:figure4}
\end{figure}

To evaluate the overall performance of our solid immersion metalens, we simulated beam propagation from the vertical and horizontal interfacial dipoles through to form an image by an `ideal' tube lens. We assume the metalens is placed at $z=0$ with the dielectric substrate ranging from z = -100 $\mu{m}$ to z = 0 $\mu{m}$ and z $>$ 0 being vacuum. Consider first, interfacial dipoles on the metalens's optical axis at (0, 0, -100 $\mu{m}$). Figures 4a, 4b and 4c present the beam propagation before and after the metalens for the vertical dipole at $\phi$ = 0, the horizontal dipole at $\phi$ = 0 and the horizontal dipole at $\phi$ = $\frac{\pi}{2}$, respectively. The phase profile just after the lens is indeed constant showing effective collimation by the metalens. Propagation beyond the metalens considers the angular spectrum of plane waves at $z=0$ and also show a flat phase profile at the field angle of $0^o$. 

In order to show the effect of aberrations at larger field angles, we place a tube lens at z = 20 $\mu{m}$. The focal length of the tube lens is set to be 5000 $\mu{m}$, so that the magnification of the whole optical system is $108 \times$. The tube lens is assumed to be perfect and thus introduces no aberrations. We consider four more vertical or horizontal interfacial dipoles at the positions (0, $\pm{1.75}$ $\mu{m}$, -100 $\mu{m}$) and (0, $\pm{3.49}$ $\mu{m}$, -100 $\mu{m}$). These dipoles correspond to the field angles $\theta_f=$ $\pm{1}^o$ and $\pm{2}^o$. For the interfacial dipoles displaced from the optical axis along the $y$-direction, the refracted angle just after the metalens can be derived by the equation of the generalized laws of reflection and refraction:\cite{yu2011light}
\begin{equation}
\sin\theta_t-n_S\sin\theta_i=\frac{\lambda_0}{2\pi}\frac{d\phi}{dy}
\end{equation}
\noindent where $\lambda_0$ is the wavelength of the incident light in vacuum, $\theta_i$  is the incident angle in the substrate, $\theta_t$ is the refracted angle in vacuum, and $d\phi$ is the phase difference between two points separated by a distance of $dy$ at the interface. The calculated refracted angle $\theta_t$ can then be used to predict beam propagation beyond the lens. We have simulated the imaging of vertical and horizontal interfacial dipole intensity distributions, as a function of position along the $y$ axis, in the focal plane of the tube lens, as shown in Figure 4d and 4e, respectively. The intensity distributions for both dipole orientations at $0^o$ field angle are diffraction limited Airy disks. However, for the field angles of $\pm{1}^o$ and $\pm{2}^o$, significant aberration occurs. 

The displacement of the central peaks from the optical axis in the focal plane of the tube lens are calculated by the product of the displacement of the dipole source and the magnification of the optical system. The displacements of the central peaks are $\pm{375.9}$ $\mu{m}$ and $\pm{187.9}$ $\mu{m}$ for the field angles of $\pm{2}^o$ and $\pm{1}^o$, respectively. From the simulation, the largest peaks occur at $\pm{351.6}$ $\mu{m}$ and $\pm{169.7}$ $\mu{m}$ for the field angles of $\pm{2}^0$ and $\pm{1}^o$, which are in good agreement with the theoretical expectations. The series of peaks appearing at smaller displacement are attributed to lens aberration, in particular coma.\cite{groever2017meta} Aberrations can be reduced by using a smaller NA. For example, we have simulated the intensity distributions of five vertical interfacial dipoles with a lower NA = 0.89 metalens ($90$ $\mu{m}$ in diameter), as shown in Figure 4f. We can see the intensity distributions for the field angles of $\pm{2}^o$ and $\pm{1}^o$ become more Airy disk-like. While the lower NA successfully reduces the coma aberration, this is unacceptable here as it drastically reduces the collection efficiency. 

\begin{figure}[t]
\centering
\includegraphics[width=0.6\linewidth]{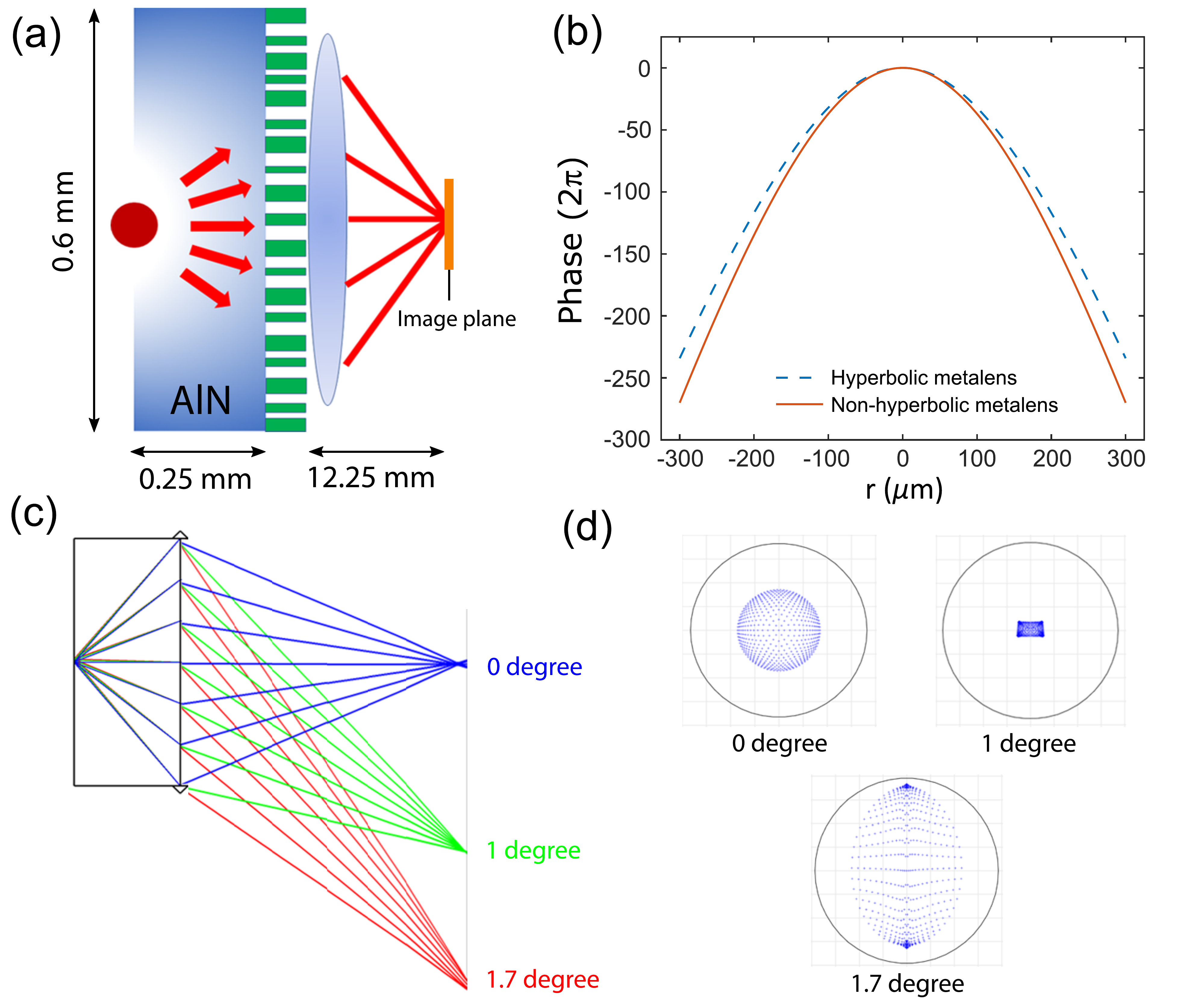}
\caption{The design of singlet solid immersion metalens for abberation correction. (a) Schematic of singlet metalens. The red circle represent the interfacial dipole. The blue rectangle regime is the substrate. The green part denote the metasurface. The red lines represent the light beam. The materials of the substrate and nano pillors are AlN and TiO$_2$ respectively. (b) The phase profiles of the singlet non-hyperbolic metasurface and a normal hyperbolic metasurface. (c) The ray tracing path diagram of the system with the incident condition of 0, 1, and 1.7 degrees field angle. (d) Spot diagram of the singlet metalens system. The blue dots indicate the image of interfacial dipole. The circle outside the blue dots denotes the Airy disk, with the radius of 14.6 $\mu{m}$.}
\label{fig:figure5}
\end{figure}

In order to correct the abberation of the metalens for a range of field angles, we have designed a singlet metalens with a non-hyperbolic phase profile. The schematic of the system is shown in Figure 5a. The substrate is chosen to be AlN with a refractive index of 2.15. In our simulation, the thickness of the AlN substrate and focal length of the metalens are 0.25 mm. The radius of the meatasurface is 0.3 mm. So the numerical aperture of the metalens is 1.65. The focal length of the tubelens was chosen to be 12.25 mm, so the objects can be magnified by $105\times$. The phase profile of the singlet metalens is described by:
\begin{equation}
    \phi_{target}=-\frac{2\pi}{\lambda}n_S\left[\sqrt{r^2+f^2}-f\right]+\sum_{m=1}^{5} a_m(\frac{r}{R_s})^{2m}
\end{equation}
where $\lambda$ is the emission wavelength of the dipole (600 nm), $n_S$ is the refractive index of substrate, $f$ is the focal distance (0.25 mm), $r$ is the radial coordinate, $R_s=300$ $\mu$m is the normal radius, $a_m$ are the optimization parameters. The optimised values, $a_1=-46.09$, $a_2=16.46$, $a_3=-10.12$, $a_4=5.124$, and $a_5=-1.281$.

The phase profiles of our designed singlet non-hyperbolic metalens and the normal hyperbolic metalens are shown in Figure 5b. Compared to the normal hyperbolic lens, the additional terms increase the phase gradient at the edge of the metalens. Figure 5c reveals the ray-tracing optical path diagram of this system (Zemax). Here, the vertical double-arrow line represents the metalens and the paraxial tube lens. The distance between them is zero. In Zemax, we have used a Binary 2-surface to act as a metalens and choose three field angles ($0^o$, $1^o$, $1.7^o$) to run the optimization. The corresponding results are shown in Figure 5d. The information of the spot sizes and the positions are shown in the Table \ref{tab:singlet metalens}. The black circle in Figure 5d denotes the Airy disk with a radius of 14.6 $\mu$m. As we can see in Figure 5d and the Table \ref{tab:singlet metalens}, all the root means square (RMS) spot radii of the incident angle are inside the Airy disk circle. This means we can achieve $\pm{1.7}^o$ field angle corresponding to $14.8$ $\mu$m field of view with negligible image aberration. 

\begin{table}[t]
\centering
\caption{\bf Spot size and the position of different field angles}
\begin{tabular}{c|cccc}
\hline
Field angle /degrees & 0 & 1 & 1.7 \\

\hline
Source position /$\mu{m}$ & 0 & 4.36 & 7.42 \\

\hline
Image position /$\mu{m}$ & 0 & -460.2 & -784 \\

\hline
RMS Spot Radius /$\mu{m}$ & 5.76 & 1.86 & 10.6 \\

\hline
Max Spot Radius /$\mu{m}$ & 6.82 & 2.49 & 14.6 \\
\hline
\end{tabular}
\label{tab:singlet metalens}
\end{table}

\section*{Conclusion}
In this paper, we have reported an efficient and compact solid immersion metalens to collect emission from dipole emission centres. Dipoles are placed at the dielectric interface of vacuum and substrate so that most of the dipole energy is directed into the substrate. A large proportion of this energy is then collected by a high NA metalens on the opposite side of the substrate. The thickness of the whole device is no more than 250 $\mu{m}$, and may be varied to control the NA. According to the simulation result, the collection efficiency of the collection device is $>85$\%, for both dipole orientations. While aberrations are pervasive for non-zero field angles, mitigation using a non-hyperbolic gradient phase metalens is shown to be quite effective. We thus report a high numerical aperture (NA=1.65), $105\times$ magnification, solid immersion singlet metalens system with a low aberration over a ${14.8}$ $\mu$m field of view. 

\section*{Acknowledgments}
This  project  has  received  funding  from  the  European  Union’s  Horizon  2020 research  and  innovation  programme under  the  Marie  Skłodowska-Curie  grant agreement No 844591. The authors thank M$\acute{o}$nica Mota for technical support.

\section*{Disclosures}
The authors declare no conflicts of interest.

\bibliography{sample}

\end{document}